\documentclass[preprintnumbers, showpacs, floatfix,twocolumn,preprintnumbers, letterpaper, superscriptaddress,nofootinbib]{revtex4}
\usepackage{amsfonts}
\usepackage{amsmath}
\usepackage{amssymb,epsf}
\usepackage{latexsym}
\usepackage{graphicx,epsfig}
\usepackage{amssymb}
\usepackage{float}
\usepackage{subfigure}
\usepackage{epstopdf}
\usepackage[colorlinks=true,citecolor=blue,linkcolor=blue,urlcolor=black]{hyperref}
\usepackage{dcolumn}
\usepackage{psfrag}
\usepackage{wrapfig}
\usepackage{makeidx}
\usepackage{epsf}

\graphicspath{{Images/}}

\begin{document}

\title{Microscopic Origin of Black Hole Reentrant Phase Transitions}
\author{M. Kord Zangeneh}
\email{kordzangeneh.mehdi@gmail.com}
\affiliation{Physics Department, Faculty of Science, Shahid Chamran University of Ahvaz,
Ahvaz 61357-43135, Iran}
\affiliation{Research Institute for Astronomy and Astrophysics of Maragha
(RIAAM)-Maragha, IRAN, P. O. Box: 55134-441}
\author{A. Dehyadegari}
\email{adehyadegari@shirazu.ac.ir}
\affiliation{Physics Department and Biruni Observatory, Shiraz University, Shiraz 71454,
Iran}
\author{A. Sheykhi}
\email{asheykhi@shirazu.ac.ir}
\affiliation{Research Institute for Astronomy and Astrophysics of Maragha
(RIAAM)-Maragha, IRAN, P. O. Box: 55134-441}
\affiliation{Physics Department and Biruni Observatory, Shiraz University, Shiraz 71454,
Iran}
\author{R. B. Mann}
\email{rbmann@uwaterloo.ca}
\affiliation{Department of Physics and Astronomy, University of Waterloo, Waterloo,
Ontario, Canada, N2L 3G1}

\begin{abstract}
Understanding the microscopic behavior of the black holes ingredients has
been one of the important challenges in black holes physics during the past
decades. In order to shed some light on the microscopic structure of black
holes, in this paper, we explore a recently observed phenomenon for black
holes namely reentrant phase transition, by employing the Ruppeiner
geometry. Interestingly enough, we observe two properties for the phase
behaviour of small black holes that leads to reentrant phase transition.
They are correlated and they are of the interaction type. For the range of
pressure in which the system underlies reentrant phase transition, it
transits from large black holes phase to small one which possesses higher
correlation than the other ranges of pressures. On the other hand, the type
of interaction between small black holes near large/small transition line,
differs for usual and reentrant phase transitions. Indeed, for usual case,
the dominant interaction is repulsive whereas for reentrant case we
encounter with an attractive interaction. We show that in reentrant phase
transition case, the small black holes behave like a Bosonic gas whereas in
the usual phase transition case, they behave like a quantum anyon gas.
\end{abstract}

\maketitle

\section{Introduction}

A reentrant phase transition (RPT) occurs when a monotonic variation of any
thermodynamic quantity gives rise to more than one phase transitions (PTs)
such that the initial and final states are macroscopically the same. This
phenomenon was first discovered in the nicotine/water mixture during a
procedure in which, by increasing the temperature at a sufficient fixed
percentage of nicotine, the homogeneous mixed state separated into distinct
nicotine/water phases and then the homogeneous state reappeared \cite{Hud}.
More often, as a result of two (or more) competing driving mechanisms, such
behavior has also been observed in multicomponent fluid systems, gels,
ferroelectrics, liquid crystals, and binary gases as well as non-commutative
spacetimes \cite{Pan} (for more details, see the review \cite{Nar}).

In the context of black hole (BH) physics, RPT has been first discovered for
four-dimensional Born-Infeld (BI) charged anti-de Sitter (AdS) BHs \cite%
{1208.6251}. In this case, for a certain range of pressures, when
temperature is lowered monotonically, a large/small/large BHs reentrant
phase transition occurred. Further studies show that for higher than
four-dimensional BI-AdS BHs, there is no RPT \cite{1311.7299}. In \cite%
{1306.5756}, $d$-dimensional singly-spinning Kerr-AdS BHs were studied and
it was shown that RPT appears for $d\geq 6$. Remarkably, in these two BH
systems, an RPT is accompanied by Hawking-Page (HP) phase transition. This
fact is interesting and important since it has been shown that the HP phase
transition is related to a confinement/deconfinment PT in quark-gluon plasma
\cite{witten}. More studies on RPT in higher-dimensional single- and
multi-spinning Kerr-AdS and Kerr-dS BHs have been carried out in \cite%
{1308.2672,1401.2586,1507.08630,1510.00085}. It is worth mentioning that
these examples of BH RPTs are accompanied by a jump at the global minimum of
the Gibbs free energy. This discontinuity is referred to as zeroth-order PT
and seen for instance in superfluidity and superconductivity \cite{Mas}.
Recently, it has been shown that the zeroth-order PT can take place as well
in an extended phase space of charged dilaton black holes \cite{AAA}. RPTs
have also been observed in frameworks consisting of higher-curvature
corrections \cite{1406.7015,1402.2837,1505.05517,1412.5028,1509.06798}. In
these kinds of gravity theories it is possible to find multiple RPTs, and/or
RPTs in which there is no zeroth-order PT, with the RPT taking place by a
succession of some first order PTs \cite{1406.7015,1505.05517,1412.5028}.

In this paper, we explore a possible microscopic origin of the black hole
RPT via Ruppeiner geometry \cite{Rup1,Rup2}. We compare the behaviors of
Ricci Scalar of Ruppeiner geometry $R$ (refereed to as Ruppeiner invariant)
for the situations in which an RPT appears and try to infer its microscopic
origin. We do this in the case of BI-AdS BHs as well as singly-spinning
Kerr-AdS BHs. The sign of the Ruppeiner invariant indicates the dominant
interaction between possible molecules of a BH ($R>0$: repulsion, $R<0$:
attraction and $R=0$: no interaction) \cite{Rup3,Rup4,Rup5}, while its
magnitude is a measure of the average number Planck areas on the event
horizon that are correlated with each other \cite{Rup6} (for more
information, see \cite{Rup7,Rup8} and references therein). We note that the
microscopic behavior of possible BH molecules has been previously studied
via Ruppeiner geometry \cite{1502.00386,1602.03711,1607.05333,1610.06352}.

Our paper is organized as follows. In section \ref{secII}, we review the
subject of RPT in the context of black hole thermodynamics. For this
purpose, we shall consider two cases: static AdS black holes in the presence
of nonlinear BI electrodynamics and spinning Kerr-AdS black holes. We shall
use the terminology SPT (for standard phase transition) to denote any phase
transition that is not reentrant (i.e. not RPT), whereas PT shall refer to
any possible phase transition without distinction. In section \ref{RPT}, we
study the Ruppeiner geometry of SPTs and RPTs to understand the microscopic
origin of the latter. In section \ref{conclusion}, we summarize the results
we found in this paper.

\section{Review of black hole RPT\label{secII}}

In this section, we review thermodynamics of higher-dimensional BI-AdS BHs
as well as singly-spinning Kerr-AdS BHs and discuss the situations under
which the RPT appears in these configurations. Our discussions here are
based on Refs. \cite{1311.7299} and \cite{1306.5756}.

\subsection{BI-AdS black holes}

\begin{figure*}[t]
\centering{%
\subfigure[~$G-T$ diagram for different values of $P$.
Solid blue and dashed red lines correspond to
$C_{P}>0$ and $C_{P}<0$, respectively.]{
   \label{fig1a}\includegraphics[width=.46\textwidth]{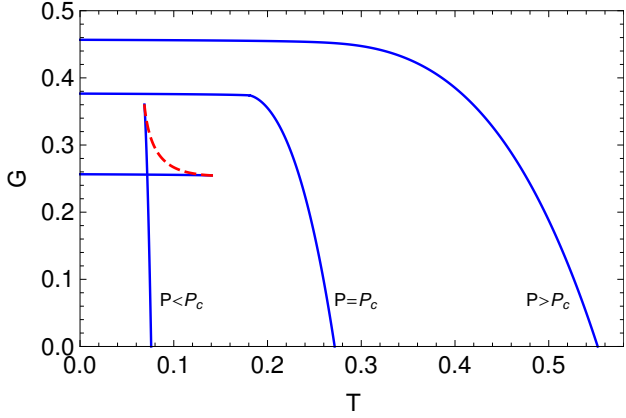}\qquad}}
\subfigure[~$P-T$ diagram. $c$ denotes the critical point.]{
   \label{fig1b}\includegraphics[width=.46\textwidth]{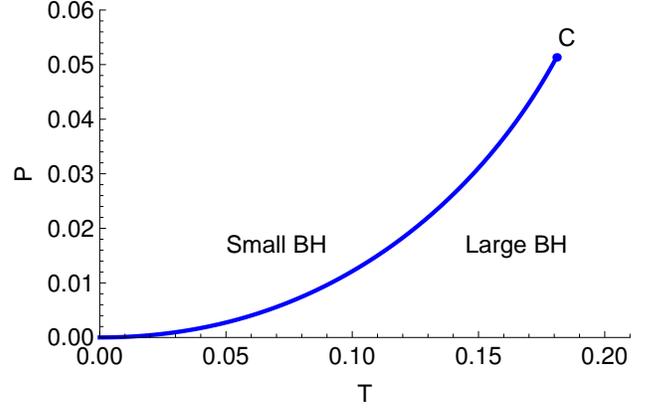}}
\caption{The behavior of $G$ and $P$ versus $T$ for $5$-dimensional BI-AdS
BHs with $Q=1$ and $b=0.035$ where $(P_{c},T_{c})=(0.051,0.181)$. This
behavior is the same for $(n\geq 5)$-dimensional BI-AdS and $4$- and $5$%
-dimensional Kerr-AdS BHs.}
\label{fig1}
\end{figure*}

The metric of $d$-dimensional BI-AdS BH is%
\begin{equation}
ds^{2}=-f(r)dt^{2}+\frac{dr^{2}}{f(r)}+r^{2}d\Omega _{d-2}^{2},
\end{equation}%
in which $d\Omega _{d-2}^{2}$ is the line element of ($d-2$)-dimensional
hypersurface with constant scalar curvature $(d-2)(d-3)$ and volume $\omega $
and%
\begin{eqnarray}
f(r) &=&1+\frac{r^{2}}{l^{2}}-\frac{m}{r^{d-3}}  \notag \\
&&+\frac{4b^{2}r^{2}}{(d-1)(d-2)}\left( 1-\sqrt{1+\frac{(d-2)(d-3)q^{2}}{%
2b^{2}r^{2d-4}}}\right)  \notag \\
&&+\frac{2(d-2)q^{2}}{(d-1)r^{2d-4}}  \notag \\
&&\times {_{2}}F_{1}\left[ \frac{d-3}{2d-4},\frac{1}{2},\frac{3d-7}{2d-4},-%
\frac{(d-2)(d-3)q^{2}}{2b^{2}r^{2d-4}}\right] ,  \notag \\
&&
\end{eqnarray}%
where $l$ is the AdS radius of spacetime, $m$ and $q$ are related to total
mass and charge of BH as%
\begin{equation}
Q=\frac{q\omega }{4\pi }\sqrt{\frac{(d-2)(d-3)}{2}}\text{, \ and \ }M=\frac{%
(d-2)m\omega }{16\pi },
\end{equation}%
where $\omega$ represents the area of the unit $(d-2)$-sphere. Using the
fact that $f(r_{+})=0$ where $r_{+}$ is the radius of outermost horizon, we
can find the constant $m$ in terms of $r_{+}$ and express the total mass of
BH as function of $r_{+}$,
\begin{eqnarray}
M &=&\frac{(d-2)\omega r_{+}^{d-3}}{16\pi }\left[ 1+\frac{r_{+}^{2}}{l^{2}}%
\right.  \notag \\
&&\left. +\frac{4b^{2}r_{+}^{2}}{(d-1)(d-2)}\left( 1-\sqrt{1+\frac{16\pi
^{2}Q^{2}}{b^{2}\omega ^{2}r_{+}^{2d-4}}}\right) \right.  \notag \\
&&\left. +\frac{64\pi ^{2}Q^{2}}{(d-1)(d-3)\omega ^{2}r_{+}^{2d-6}}\right.
\notag  \label{eq:8a} \\
&&\left. \times {_{2}}F_{1}\left[ \frac{d-3}{2d-4},\frac{1}{2},\frac{3d-7}{%
2d-4},-\frac{16\pi ^{2}Q^{2}}{b^{2}\omega ^{2}r_{+}^{2d-4}}\right] \right] .
\end{eqnarray}%
In the above expression $b$ is the nonlinear BI parameter appears in the BI
Lagrangian, $L_{BI}=4b^{2}(1-\sqrt{1+F^{\mu \nu }F_{\mu \nu }/2b^{2}})$. The
electrodynamics field tensor $F_{\mu \nu }=\partial _{\lbrack \mu }A_{\nu ]}$
and gauge potential $A_{\nu }=\left( A_{t},0,0,\cdots \right) $ is given by
\begin{eqnarray}
A_{t} &=&-\frac{q}{r^{d-3}}\sqrt{\frac{d-2}{2(d-3)}}  \notag \\
&&\times {_{2}}F_{1}\left[ \frac{d-3}{2d-4},\frac{1}{2},\frac{3d-7}{2d-4},-%
\frac{(d-2)(d-3)q^{2}}{2b^{2}r^{2d-4}}\right] .  \notag \\
&&
\end{eqnarray}%
In the limiting case where $b\rightarrow\infty$, the BI Lagrangian
reproduces the linear Maxwell one. The electromagnetic potential defined by%
\begin{equation}
\Phi =A_{\mu }\chi ^{\mu }\left\vert _{r\rightarrow \infty }-A_{\mu }\chi
^{\mu }\right\vert _{r=r_{+}},
\end{equation}%
where $\chi =\partial _{t}$ is the Killing vector, can also be calculated as%
\begin{eqnarray}
\Phi &=&\frac{4\pi Q}{(d-3)\omega r_{+}^{d-3}}  \notag \\
&&\times {_{2}}F_{1}\left[ \frac{d-3}{2d-4},\frac{1}{2},\frac{3d-7}{2d-4},-%
\frac{16\pi ^{2}Q^{2}}{b^{2}\omega ^{2}r_{+}^{2d-4}}\right] .
\end{eqnarray}%
The cosmological constant is fixed as $\Lambda =-(d-1)(d-2)/2l^{2}$ and
related to the thermodynamic pressure as (in Planck unit) \cite%
{CaldarelliEtal:2000, KastorEtal:2009, CveticEtal:2010, Dolan:2012}%
\begin{equation}
P=-\frac{\Lambda }{8\pi }=\frac{(d-1)(d-2)}{16\pi l^{2}}.  \label{Pressure}
\end{equation}%
The corresponding Hawking temperature associated with the event horizon $%
r_{+}$, and the entropy of BH are given by
\begin{eqnarray}
T &=&\frac{1}{4\pi }\left[ \frac{(d-1)r_{+}}{l^{2}}+\frac{(d-3)}{r_{+}}%
\right.  \notag \\
&&\left. +\frac{4b^{2}r_{+}}{d-2}\left( 1-\sqrt{1+\frac{16\pi ^{2}Q^{2}}{%
b^{2}\omega ^{2}r_{+}^{2d-4}}}\right) \right] ,  \label{TempBI} \\
S &=&\frac{\omega r_{+}^{d-2}}{4}.  \label{eq:10a}
\end{eqnarray}%
\begin{figure*}[t]
\subcapraggedrighttrue\centering{%
\subfigure[~The behavior of $G$ for different pressures where
$(P_c,P_z,P_t)=(0.0957,0.0578,0.0553)$ and $(T_c,T_z,T_t)=(0.3002,0.2349,0.2331)$. \newline
Note that various curves are shifted for clarity.]{
   \label{fig2a}\includegraphics[width=.46\textwidth]{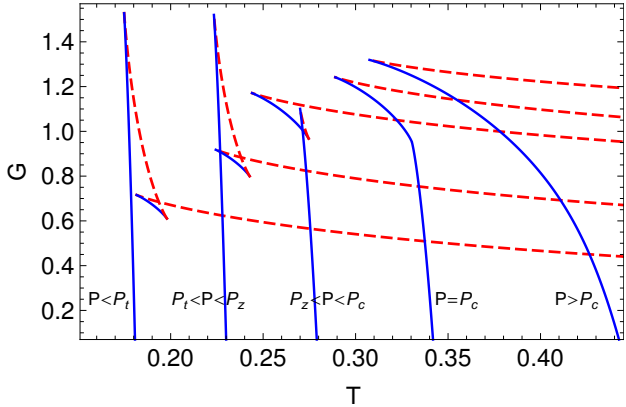}\qquad}}
\subfigure[~The behavior of $G$ for $P=0.0568$. The arrows show the direction of event
horizon radius increase. The zeroth-order PT takes place at $T_0=0.2335$.]{
   \label{fig2b}\includegraphics[width=.46\textwidth]{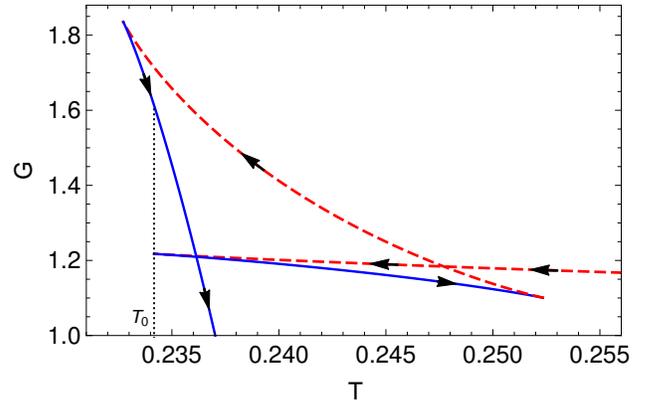}}
\caption{The behavior of $G$ versus $T$ for $6$-dimensional Kerr-AdS black
hole with $J=1$. Solid blue and dashed red lines are corresponding
respectively to $C_{P}>0$ and $C_{P}<0$. This behavior is qualitatively the
same for $4$-dimensional BI-AdS and $(n\geq 6)$-dimensional Kerr-AdS BHs.}
\label{fig2}
\end{figure*}

\begin{figure}[t]
\epsfxsize=8cm \centerline{\epsffile{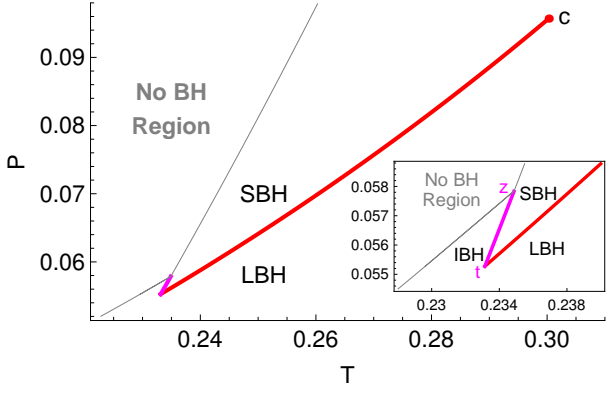}}
\caption{Phase diagram for $6$-dimensional Kerr-AdS BH with $J=1$. $c$
denotes the critical point. We zoom in on RPT region to make it more clear.
NO BH REGION corresponds to a forbidden region of parameter space. This
behavior is the same for $(n\geq 6)$-dimensional Kerr-AdS and $4$%
-dimensional BI-AdS BHs.}
\label{fig3}
\end{figure}
Interpreting the mass of BH, $M$, as the enthalpy rather than the internal
energy of the gravitational system \cite{KastorEtal:2009}, we can write the
first law of thermodynamics as
\begin{equation}
dM=TdS+\Phi dQ+VdP+\mathfrak{B}db,
\end{equation}%
in which $M$ is a function of BI coupling coefficient as well as other usual
thermodynamic parameters \cite{Gunasekaran:2012dq}. The conjugate quantity
of $b$ is given by
\begin{eqnarray}
\mathfrak{B} &=&\frac{\omega br_{+}^{d-1}}{2(d-1)\pi }\left( 1-\sqrt{1+\frac{%
16\pi ^{2}Q^{2}}{b^{2}\omega ^{2}r_{+}^{2d-4}}}\right)  \notag \\
&&+\frac{4Q^{2}\pi }{(d-1)b\omega r_{+}^{d-3}}  \notag  \label{eq:21a} \\
&&\times {_{2}}F_{1}\left[ \frac{d-3}{2d-4},\frac{1}{2},\frac{3d-7}{2d-4},-%
\frac{16\pi ^{2}Q^{2}}{b^{2}\omega ^{2}r_{+}^{2d-4}}\right] .
\end{eqnarray}%
which is referred to as `BI vacuum polarization'. Also, the thermodynamic
volume conjugate to the pressure is $V=\omega r_{+}^{d-1}/\left( d-1\right) $%
. The generalized Smarr formula in the extended phase space can be obtained
by scaling argument as
\begin{equation}
M=\frac{d-2}{d-3}TS+\Phi Q-\frac{2}{d-3}VP-\frac{1}{d-3}\mathfrak{B}b.
\end{equation}%
To examine the PT behaviors (or thermodynamic behaviors), in addition to the
equation of state $T=T(P,V)$, which can be calculated by eliminating $l^{2}$
between (\ref{Pressure}) and (\ref{TempBI}), we study the Gibbs free energy $%
G=M-TS$,%
\begin{eqnarray}
G(T,P) &=&\frac{\omega }{16\pi }\left[ r_{+}^{d-3}-\frac{16\pi Pr_{+}^{d-1}}{%
(d-1)(d-2)}\right.  \notag \\
&&\left. -\frac{4b^{2}r_{+}^{d-1}}{(d-1)(d-2)}\left( 1-\sqrt{1+\frac{16\pi
^{2}Q^{2}}{b^{2}\omega ^{2}r_{+}^{2d-4}}}\right) \right.  \notag \\
&&\left. +\frac{64(d-2)^{2}\pi ^{2}Q^{2}}{(d-1)(d-3)\omega ^{2}r_{+}^{d-3}}%
\right.  \notag \\
&&\left. \times {_{2}}F_{1}\left[ \frac{d-3}{2d-4},\frac{1}{2},\frac{3d-7}{%
2d-4},-\frac{16\pi ^{2}Q^{2}}{b^{2}\omega ^{2}r_{+}^{2d-4}}\right] \right] .
\notag  \label{eq:32a} \\
&&
\end{eqnarray}%
Note that in our study on BI-AdS BH, we treat $Q$ and $b$ as fixed
variables.
\begin{figure*}[t]
\centering{%
\subfigure[~$6$-dimensional Kerr-AdS with $J=1$.]{
   \label{fig4a}\includegraphics[width=.46\textwidth]{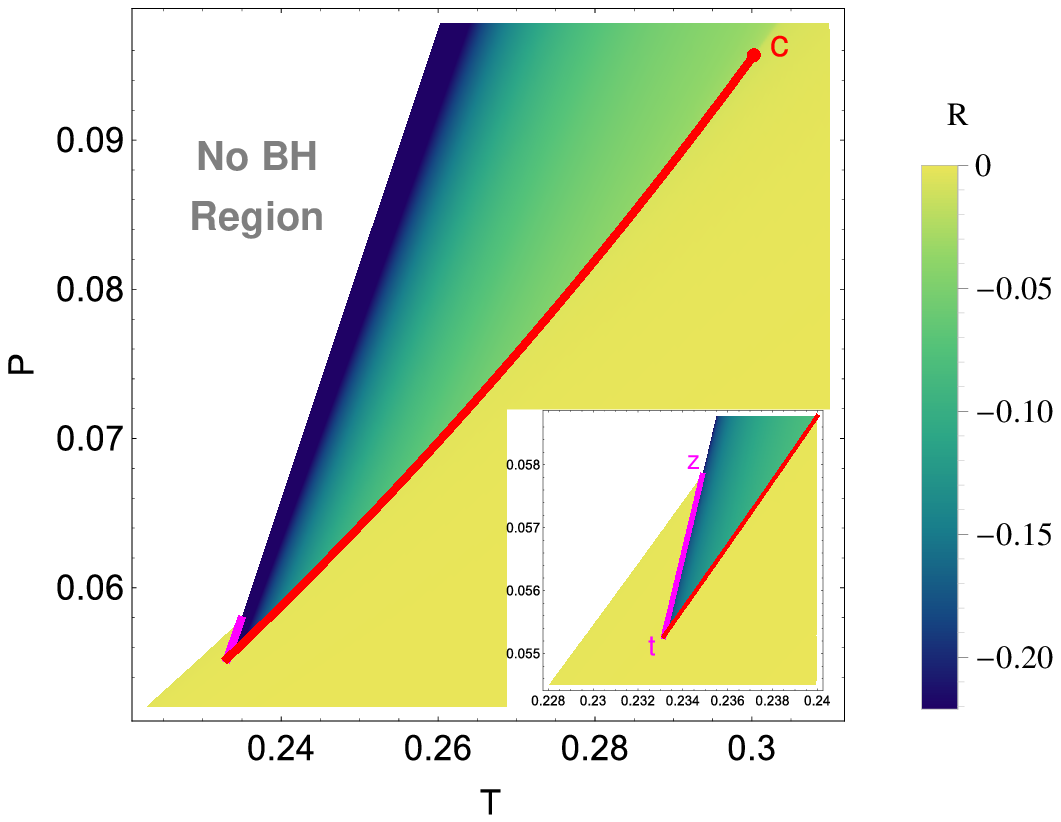}\qquad}}
\subfigure[~$4$-dimensional BI-AdS with $Q=1$ and $b=0.035$.]{
   \label{fig4b}\includegraphics[width=.46\textwidth]{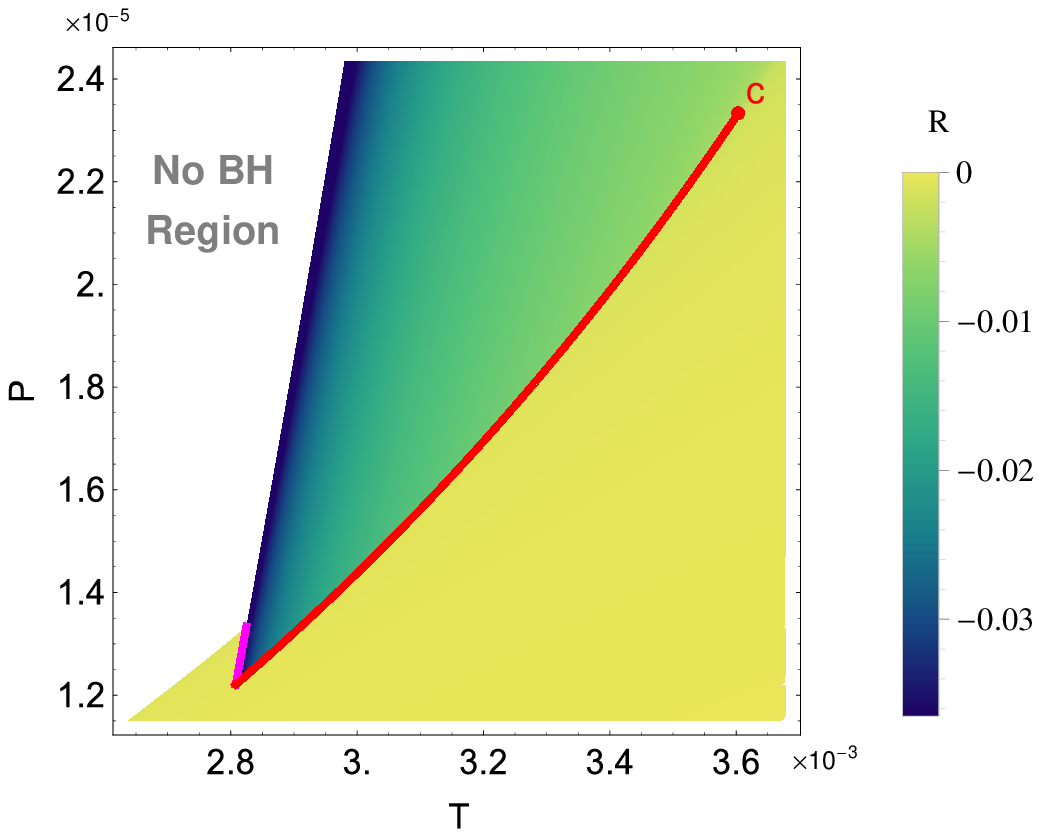}}
\caption{The $R-P-T$ diagram for $4$-dimensional BI-AdS and $6$-dimensional
Kerr-AdS BHs. This behavior is the same for $(n\geq 6)$-dimensional Kerr-AdS
and $4$-dimensional BI-AdS BHs.}
\label{fig4}
\end{figure*}

\subsection{Singly-spinning Kerr-AdS black holes}

In $d$-dimensional spacetime the metric of singly spinning Kerr-AdS black
holes may be written
\begin{eqnarray}
ds^{2} &=&-\frac{\Delta }{\rho ^{2}}(dt-\frac{a}{\Xi }\sin ^{2}\!\theta
d\varphi )^{2}+\frac{\rho ^{2}}{\Delta }dr^{2}+\frac{\rho ^{2}}{\Sigma }%
d\theta ^{2}  \notag \\
&+&\frac{\Sigma \sin ^{2}\!\theta }{\rho ^{2}}[adt-\frac{(r^{2}+a^{2})}{\Xi }%
d\varphi ]^{2}+r^{2}\cos ^{2}\!\theta d\Omega _{d-2}^{2}\,,\qquad
\end{eqnarray}%
where%
\begin{gather}
\Delta =(r^{2}+a^{2})(1+\frac{r^{2}}{l^{2}})-\frac{2m}{r^{d-5}}\text{, \ }%
\Xi =1-\frac{a^{2}}{l^{2}}\,\text{,}  \notag \\
\Sigma =1-\frac{a^{2}}{l^{2}}\cos ^{2}\!\theta \,\ \text{and \ }\rho
^{2}=r^{2}+a^{2}\cos ^{2}\!\theta \,\text{.}
\end{gather}%
The associated thermodynamic quantities read (in Planck units)%
\begin{eqnarray}
M &=&\frac{\omega }{4\pi }\frac{m}{\Xi ^{2}}\left( 1+\frac{(d-4)\Xi }{2}%
\right) \,,  \label{BHOM} \\
J &=&\frac{\omega }{4\pi }\frac{ma}{\Xi ^{2}}\,,  \label{BHJ} \\
{\Omega }_{H} &=&\frac{a}{l^{2}}\frac{r_{+}^{2}+l^{2}}{r_{+}^{2}+a^{2}}\,, \\
P &=&\frac{-\Lambda }{8\pi }=\frac{(d-1)(d-2)}{16\pi l^{2}}\,,\quad \\
T &=&\frac{1}{2\pi }\left[ \left( 1+\frac{r_{+}^{2}}{l^{2}}\right) \left(
\frac{r_{+}}{a^{2}+r_{+}^{2}}+\frac{d-3}{2r_{+}}\right) -\frac{1}{r_{+}}%
\right] \,,\qquad  \label{BHT} \\
S &=&\frac{\omega }{4}\frac{(a^{2}+r_{+}^{2})r_{+}^{d-4}}{\Xi }\,,
\label{BHS}
\end{eqnarray}%
where $m$ can be calculated by setting $\Delta \left( r_{+}\right) =0$. We
find
\begin{equation}
m=\left( r_{+}^{2}+a^{2}\right) \left( \frac{r_{+}^{d-5}}{2}+\frac{%
r_{+}^{d-3}}{2l^{2}}\right) .
\end{equation}%
One can also find the first law of BH thermodynamics and Smarr formula as
\cite{KastorEtal:2009}%
\begin{eqnarray}
dM &=&TdS+\Omega _{H}dJ+VdP\,,  \label{1st} \\
\frac{d-3}{d-2}M &=&TS+\Omega _{H}J-\frac{2}{d-2}VP\,,  \label{Smarr}
\end{eqnarray}%
where the volume conjugate to pressure is%
\begin{equation}
V=\frac{\omega r_{+}^{d-1}}{d-1}\left[ 1+\frac{a^{2}}{\Xi }\frac{%
1+r_{+}^{2}/l^{2}}{(d-2)r_{+}^{2}}\right] \,.  \label{VBH}
\end{equation}%
The Gibbs free energy $G=M-TS$ governing the thermodynamic behavior of the
system reads%
\begin{eqnarray}
G\left( T,P,J\right) \! &=&\!\frac{\omega r_{+}^{d-5}}{16\pi \Xi ^{2}}
\notag \\
&&\times \left( 3a^{2}+r_{+}^{2}-\frac{(r_{+}^{2}\!-\!a^{2})^{2}}{l^{2}}+%
\frac{3a^{2}r_{+}^{4}+a^{4}r_{+}^{2}}{l^{4}}\right) .  \notag \\
&&
\end{eqnarray}%
Eliminating $m$, $a$ and $r_{+}$ in Eqs. (\ref{BHOM})-(\ref{BHS}) in terms
of the basic thermodynamic variables, one can numerically obtain the
equation of state $T=T(P,V)$\textbf{\ } in the canonical ensemble (with
fixed $J$).

\subsection{Reentrant phase transition}

The thermal stability of a thermodynamic system may be determined by the
sign of system's response functions. One of theses response functions is
specific heat at constant pressure. For the BHs under consideration, the
specific heat is given by
\begin{equation}
C_{P}=T\left( \frac{\partial S}{\partial T}\right) _{P,\{Y\}} \text{ \ \ \ \
}\{Y\}=\left\{
\begin{array}{ll}
b, \text{ }Q & \text{for BI-AdS} \\
J & \text{for Kerr-AdS}%
\end{array}%
\right.
\end{equation}%
Positivity (negativity) of $C_{P}$ determines the local stability
(instability) which is shown by solid blue (dashed red) lines in a $G-T$
diagram (see Figs. \ref{fig1a} and \ref{fig2}). In order to examine the
precise behavior of the system, we investigate the behavior of the Gibbs
free energy $G$. For $5$-dimensional BI-AdS black holes this is plotted in
Fig. \ref{fig1a}. From this figure, we see that for $P>P_{c}$, $G$ has
smooth behavior as a function of $T$, whereas for $P<P_{c}$ it exhibits
multi-valued behavior related to the LBH/SBH (large black hole/small black
hole) first order SPT that is similar to the SPT of the Van der Waals fluid
system. The corresponding phase diagram is shown in Fig. \ref{fig1b} in
which for $P<P_{c}$ and $T<T_{c}$, large and small BH phases are separated
by the transition line. This kind of behavior is observed for $(d\geq 5)$%
-dimensional BI-AdS BHs and also $4$- and $5$-dimensional Kerr-AdS BHs.

However for $4$-dimensional BI-AdS BHs and $(d\geq 6)$-dimensional Kerr-AdS
BHs, the thermodynamic behaviour is different. The behaviour of the Gibbs
free energy for $6$-dimensional Kerr-AdS BHs versus temperature for fixed
pressure is plotted in Fig. \ref{fig2a}. As one can see, for $P>P_{c}$ the
behavior of the system is similar to that of Schwarzschild-AdS BHs: the
upper (lower) branch corresponds to small (large) BHs with $C_{P}<0$ ($%
C_{P}>0$) and at $G(T_{HP},P_{HP})=0$, the Hawking-Page SPT occurs.
Decreasing the pressure to $P=P_{c}$, we have an additional second order PT
at $T=T_{c}$ in comparison with $P>P_{c}$ case. In the region $P_{t}<P<P_{c}$%
, a LBH/SBH first order SPT takes place which is similar to
liquid/gas Van der Waals SPT (note that $P_{t}$ shows the end of a
transition line corresponding to the first order SPT). In the
region $P_{t}<P<P_{z}$, according to the behavior of Gibbs free
energy, we have a zeroth-order PT besides the LBH/SBH first order
PT (see Fig. \ref{fig2b}). Note that $P_{z} $ determines the
starting point for observing zeroth-order PT. This kind of phase
transition takes place by a discontinuity in Gibbs free energy
that gives rise to a small/large (intermediate) BHs PT. Therefore,
generally we have a LBH/SBH/LBH phase transition -- this is an RPT
since the starting and ending phases are macroscopically similar.
For $P<P_{t}$ the situation is the same as $P>P_{c}$ case; note
that since the black hole has charge a Hawking-Page PT (which would
not conserve charge) does not take place.

The phase diagram corresponding to an RPT is depicted in the $P-T$
diagram in Fig. \ref{fig3}. The solid lines are coexistence lines
denoting the boundary of two different phases. The inset shows
that for $P>P_{z}$ there is a SPT, whereas for $P<P_{t}$ there is
no PT, even where $G$ changes sign. The RPT occurs for
$P_{t}<P<P_{z}$, where the red line signifies a first-order PT and
the magenta line signifies a zeroth-order PT. Note that there is a
forbidden region in parameter space for which no black hole
solutions exist. Similar RPT behavior takes place for $4$-dimensional BI-AdS as well as $%
(d\geq 6)$-dimensional Kerr-AdS BHs \cite{1208.6251,1306.5756}.

For $P_{t}<P=0.0567<P_{z}$ an RPT corresponding to a $6$-dimensional
Kerr-AdS BHs is depicted in Fig. \ref{fig2b}. The direction of arrows
follows increasing horizon radius. As temperature decreases the system
always moves on the lower branch in each region to maintain thermodynamic
equilibrium. Beginning on the lower steeply-sloped curve, a first order PT
takes place after which the system is on the near-flat blue curve. This is
an SPT from an LBH to an SBH. At $T=T_{0}$, a zeroth-order PT takes place at
which a finite jump in $G$ takes place and the system moves back to the
steep blue curve. This finite jump is a zeroth-order PT between an SBH and
an IBH (an LBH but of smaller radius). The entire process is an RPT.

\section{Microscopic origin of the Black Hole RPT\label{RPT}}

In this section, we will study the microscopic origin of black holes RPT by
adopting the Ruppeiner approach towards thermodynamic geometry of BHs \cite%
{Rup3,Rup4,Rup5,Rup6,Rup7,Rup8}. We define the Ruppeiner metric in $%
X^{\alpha }=\left( M,Z\right) $ space where $Z=Q$ for BI-AdS BHs and $Z=P$
for Kerr-AdS BHs. The entropy $S$ leads to the thermodynamic potential,
\begin{equation}
g_{\alpha \beta }=-\frac{\partial ^{2}S}{\partial X^{\alpha }\partial
X^{\beta }}.
\end{equation}%
The above metric can also be rewritten in the Weinhold form%
\begin{equation}
g_{\alpha \beta }=\frac{1}{T}\frac{\partial ^{2}M}{\partial Y^{\alpha
}\partial Y^{\beta }},  \label{MetGTD}
\end{equation}
where $Y^{\alpha }=(S,Z)$. The Ricci scalar corresponding to this metric is
referred to as Ruppeiner invariant $R$, and can give us some information
about the microscopic behavior of possible BH molecules. The sign of $R$
signifies the dominant interaction between microscopic constituents of a
system (the possible BH molecules in our case): positive, negative and zero
values respectively indicate repulsion, attraction, and no interaction \cite%
{Rup3,Rup4,Rup5}. Its magnitude provides a measure of the average number of
correlated constituents \cite{Rup6}; for black holes this would be the
average number of correlated Planck areas on the event horizon.

In order to study the microscopic origin of an RPT, we examine the behavior
of the Ruppeiner invariant on both sides of the coexistence lines in the $%
P-T $ diagram. We illustrate this in Fig. \ref{fig4} for both 4-dimensional
BI-AdS black holes and for 6-dimensional singly spinning Kerr-AdS
blackholes; note that both diagrams are similar to \ref{fig3}. Since there
are two PT in case of RPT, we have two transition lines. The rightmost and
leftmost transition lines are respectively related to LBH/SBH and SBH/IBH (a
smaller LBH) PTs as temperature is decreased.

We see for both cases that $R<0$, indicating that the dominant interaction
is attractive. The important point we can understand from Fig. \ref{fig4} is
that for the zeroth-order PT $R$ is much more negative in going from the SBH
to the IBH than in going from the LBH to the SBH. For the RPT, $R$ goes from
small negative (LBH) to larger negative (SBH) back to small negative (IBH)
again by decreasing the temperature at fixed pressure. According to the
definition of RPT, we know that the starting and ending phases should be the
same macroscopically. From Fig. \ref{fig4} we see that, they are almost the
same too, from a microscopic point of view, since the magnitude and sign of
the Ruppeiner invariant $R$ are the same on these phases.

\begin{figure}[h]
\epsfxsize=8.5cm \centerline{\epsffile{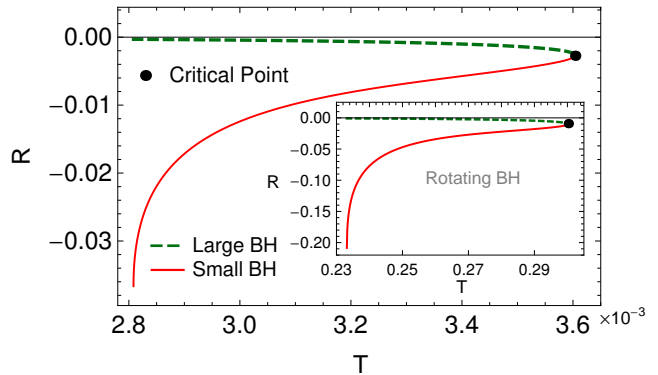}}
\caption{The behavior of the Ruppeiner invariant $R$ along the rightmost
transition line (the red line in Fig.~\protect\ref{fig4}) for $4$%
-dimensional BI-AdS and $6$-dimensional Kerr-AdS BHs. For $(n\geq 6)$%
-dimensional Kerr-AdS BHs, this behavior is qualitatively the same.}
\label{fig5}
\end{figure}
\begin{figure}[h]
\epsfxsize=8.5cm \centerline{\epsffile{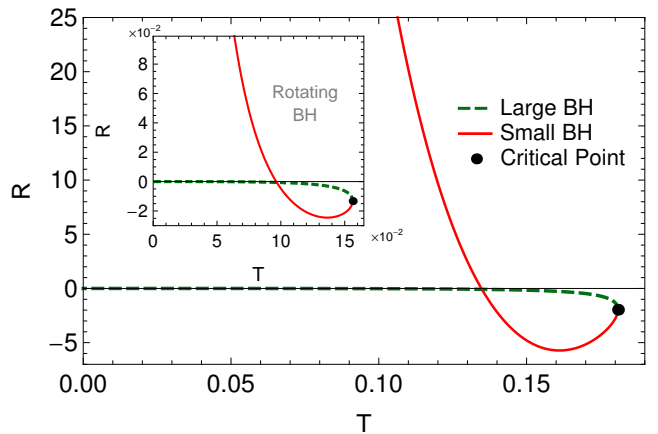}}
\caption{The behavior of Ruppeiner invariant $R$ along the transition line
\textbf{for }$5$-dimensional BI-AdS in Fig.~\protect\ref{fig1b} and Kerr-AdS
BHs with $J=1$, with the values of $b$ and $Q$ as in Fig. \protect\ref{fig1}%
. For $(n\geq 5)$-dimensional BI-AdS and $4$- and $5$-dimensional Kerr-AdS
BHs, the behavior is qualitatively the same.}
\label{fig6}
\end{figure}

We see from Fig. \ref{fig4} that when we move downward through the rightmost
transition line, the value of $R$ for points close to the line and on its
right side (i.e. the LBH phase) is near zero. This behaviour can be seen
more clearly in Fig. \ref{fig5}. For the usual first-order PT, the magnitude
of the Ruppeiner invariant $R$ for the LBH phase near the transition line is
almost the same as for the IBH i.e. it is near zero. For these black holes
their microscopic constituents attract and have weak correlation. This
behaviour is expected for large BHs where microscopic ingredients of BH are
expected to be far from each other.

However, for the SBH phase, the microscopic behaviour of BHs underlying the
usual PT and RPTs are different near the transition line. For the underlying
RPT, as we move downward near the coexistence line in Fig. \ref{fig5} on its
left side, the Ruppeiner invariant is negative and of increasing magnitude.
The dominant interaction is attractive, with the constituents much more
strongly correlated.

The behaviour of $R$ for the usual SPT in Fig. \ref{fig1b} is illustrated in
Fig. \ref{fig6}. In this case, the value $R$ near the coexistence line for
the SBH is negative at critical point but passes through zero toward
increasingly positive values (signifying increasing correlation for
repulsively interacting constituents) as the temperature decreases; the
value of $R$ is positive infinite near zero temperature. We know that for
SBHs near the transition line, the pressure is low when temperature is low,
as is clear from Fig. \ref{fig1b}. In this region of the phase diagram for
the SBH the the repulsion is high; despite this, the pressure is low.
Evidently the effects of low temperature dominate over this repulsion,
`freezing' the black hole molecules and increasing their correlation. The
LBH always has weak attraction and low correlation amongst its constituents.

It is worth noting that the RPT occurs in a special region of low pressures $%
P_{t}<P<P_{z}$, shown in Fig. \ref{fig4}. For a given pressure in this
region, by decreasing the temperature we transit from an LBH to an SBH of
greater $|R|$, indicating greater correlations in comparison to other ranges
of pressure. This high correlation becomes higher as we decrease the
temperature further in this pressure range. One of the signs of a PT is a
change from low to high correlation amongst constitutents (or vice-versa).
It seems that from a microscopic point of view, the difference making the BH
system able to undergo an RPT for this special region of pressure $%
P_{t}<P<P_{z}$ is the greater correlation between molecules of SBHs in
comparison to other ranges of pressure.

\begin{figure}[h]
\epsfxsize=8.5cm \centerline{\epsffile{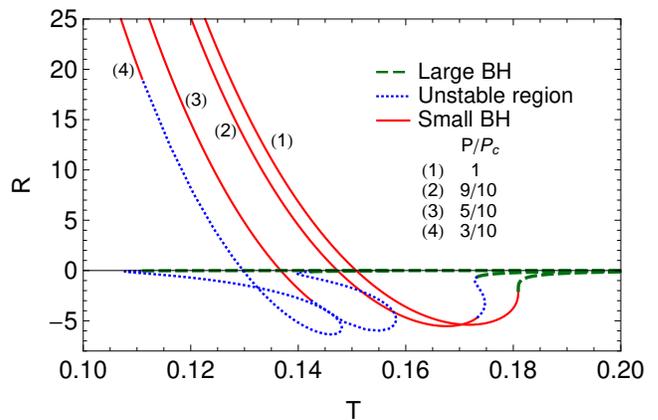}}
\caption{The behavior of Ruppeiner invariant $R$ for different given
pressures for $5$-dimensional BI-AdS BH with $Q=1$ and $b=0.035$. For other
configurations underlying usual PT, the behavior is the same. }
\label{fig7}
\end{figure}
Here, a question may arise when one compares the behaviour of the Ruppeiner
invariant $R$ for BH configurations exhibiting an RPT and ones exhibiting an
SPT. For the latter, the correlation (the magnitude of $R$) is high for
small BHs near transition line for low pressures as one can see from Fig. %
\ref{fig6} (note from Fig. \ref{fig1b} that along transition line low
temperatures lead to low pressures). Furthermore, as one can see in Fig. \ref%
{fig7}, for a given (low) pressure, when the temperature decreases and
system transits to an SBH phase, the correlation (the magnitude of $R$)
increases as temperature decreases more in this phase. The SPT/RPT
distinction is evidently one of increasing repulsion/attraction for an SBH
relative to the LBH. The attraction may help the SBH to undergo an
additional zero-order PT as temperature decreases (hence an RPT) whereas the
increasing repulsion obstructs an additional phase transitions (hence an
SPT).

As one can see in Fig. \ref{fig7}, the BH configuration with SPT shows two
different behaviors by decreasing the temperature from high to low values
for a given pressure. A part of this behavior for high temperatures carries
the Bosonic negative sign whereas another part for low temperatures looks
like a Fermionic gas. Similar behavior has been observed for a quantum gas
of anyons where the molecular volume is fixed \cite{Rup8}. The blue dotted
lines in Fig. \ref{fig7} correspond to the swallow tail part of the $G-T$
diagram for $P<P_{c}$ (Fig. \ref{fig1a}). In the swallowtail region, the
system has a thermodynamic instability on part of the curve, whereas on the
other part the specific heat $C_{P} $ at fixed $P$ (which determines the
local instability) is positive.

\section{Summary and conclusions\label{conclusion}}

In this paper, we investigated the microscopic origin of the reentrant phase
transition (RPT) in BHs thermodynamics. For this purpose, we employed the
Ruppeiner geometry towards thermodynamics geometry of BHs. We showed that
the ending and starting phases which are the same macroscopically, are the
same microscopically, too. We found out that there are no microscopic
difference between the behavior of large BHs near large-small transition
line of usual PT and RPT cases, and the main difference is in the
microscopic behavior of small black holes. We observed that for usual PT,
the dominant interaction for small BHs near transition line for low
temperatures is repulsive whereas for RPT it is attractive.

The fact that we have two phase transitions in RPT case, for the pressure in
the range of $P_{t}<P<P_{z}$, can be understood easily (see Fig. \ref{fig2a}%
). Indeed, in this region the system transits from large BH to small one
with higher correlations compared to the other regions of pressure. We
showed that with decreasing the temperature of a given pressure $P$,
correlation becomes higher and it makes the system possible to underlie
another phase transition in the region $P_{t}<P<P_{z}$.

We also explored the microscopic origin of black holes RPT by
adopting the Ruppeiner approach towards thermodynamic geometry of
BHs. It is well-known that the magnitude of the Ruppeiner
invariant $R$ measures the average number of Planck areas on the
event horizon correlated to each other or roughly speaking
correlation \cite{Rup6}. We examined the behaviour of the
Ruppeiner invariant on both sides of transition lines in $P-T$
diagram (see Fig. \ref{fig4}). We observed that in RPT case the
small BHs behave like a Bosonic gas whereas in the usual PT case
small black holes behave like a quantum anyon gas. We showed that
the freezing is more effective than the dominant interaction at
low temperature since in both the two kinds of phase transitions
(namely the usual PT and RPT) have low pressure in this region
despite different interactions.

%%%%%%%%%%%%%%%%%%%%%%%%%%%%%%%%%%%%%%%%%%%%%%%%%%%%%%%%%%%%%%%%%%%%%%%%%%%

\begin{acknowledgments}
We are grateful to Prof. Bin Wang for useful discussions and helpful
comments. AD and AS thank the research council of Shiraz University. The
work of MKZ has been supported financially by Research Institute for
Astronomy \& Astrophysics of Maragha (RIAAM) under research project No.
1/5237-57.
\end{acknowledgments}

\end{document}